\documentstyle[12pt,epsfig]{article}

\newlength{\dinwidth}
\newlength{\dinmargin}
\def\lapproxeq{\lower .7ex\hbox{$\;\stackrel{\textstyle
<}{\sim}\;$}}
\def\gapproxeq{\lower .7ex\hbox{$\;\stackrel{\textstyle
>}{\sim}\;$}}

\def\be{\begin{equation}}
\def\ee{\end{equation}}
\def\bea{\begin{eqnarray}}
\def\eea{\end{eqnarray}}

\def\pp{p\bar{p}}

\newcommand{\eq}[1]{(\ref{eq:#1})}

\begin{document}
\begin{flushright}
IPPP/03/32 \\
DCPT/03/64 \\
16 June 2003 \\

\end{flushright}

\vspace*{0.5cm}

\begin{center}
{\Large \bf Unitarity effects in hard diffraction at HERA}

\vspace*{1cm}
\textsc{A.B.~Kaidalov$^{a,b}$, V.A.~Khoze$^{a,c}$, A.D. Martin$^a$ and M.G. Ryskin$^{a,c}$} \\

\vspace*{0.5cm} $^a$ Department of Physics and Institute for
Particle Physics Phenomenology, \\
University of Durham, DH1 3LE, UK \\
$^b$ Institute of Theoretical and Experimental Physics, Moscow, 117259, Russia\\
$^c$ Petersburg Nuclear Physics Institute, Gatchina,
St.~Petersburg, 188300, Russia \\

\end{center}

\vspace*{0.5cm}

\begin{abstract}
We study how factorization breaking changes when going from diffractive deep inelastic scattering to diffractive
photoproduction of dijets. These processes offer a sensitive probe of the interplay of soft and hard mechanisms in
QCD. We demonstrate that unitarity effects are already important in the gluon distribution for $x\lapproxeq
10^{-4}$, for quite a wide range of $Q^2$.
\end{abstract}


\section{Introduction}

The investigation of hard diffractive processes initiated by real and virtual photons gives important information
on the interplay of soft and hard dynamics in QCD. These processes have been actively studied at HERA~\cite{A1}.
The hard scale is defined either by the virtuality $Q^2$ of a photon or by the $p_T(E_T)$ of jets (in diffractive
production of jets) or by the mass $M_Q$ of a heavy quark (for heavy-flavour diffractive production). For large
$Q^2$ it is possible to prove a QCD-factorization theorem~\cite{A2}, which allows one to describe the inclusive
diffractive dissociation of a photon in terms of quarks and gluons, and to predict the $Q^2$-evolution of their
distributions. Description of these processes is usually carried out in terms of the partonic distributions in the
Pomeron, and corresponds to the diagram of Fig.~1.
\begin{figure}
\begin{center}
\includegraphics[height=5cm]{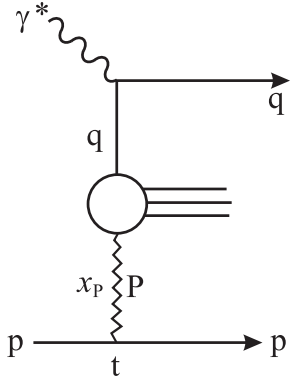}
\caption{The simplest diagram for inclusive diffractive jet production in DIS at HERA. $x_P$ is the fraction of
the longitudinal momentum of the proton carried by the Pomeron $P$.\label{fig:1}}
\end{center}
\end{figure}
In the simplest approximation the exchange of a single, factorizable Pomeron Regge-pole $P$ is assumed. We discuss
the validity of this approximation at the end of the article. On the other hand, the QCD factorization theorem is
not valid for diffractive dissociation at small $Q^2$, or for hard diffraction in hadronic interactions; see, for
example, Refs.~\cite{A2}--\cite{AB}. Rather, multi-Pomeron exchanges (of the type shown in Fig.~2(b)
\begin{figure}
\begin{center}
\includegraphics[height=7cm]{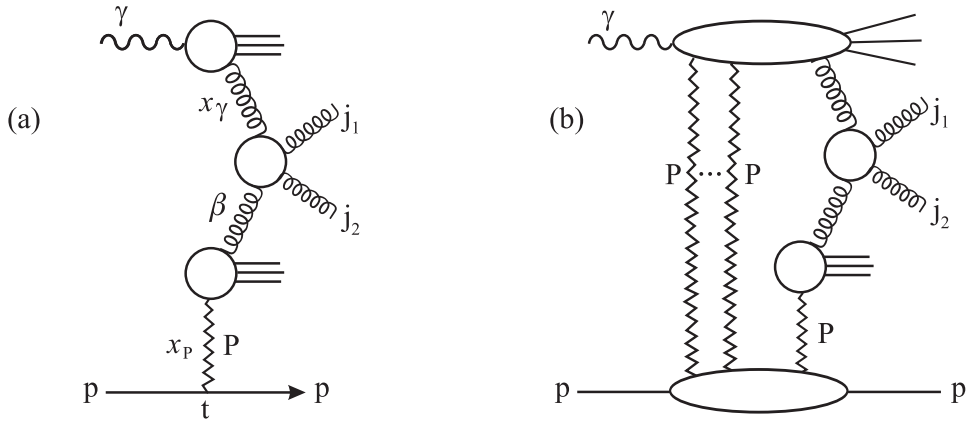}
\caption{Diffractive dijet production at HERA from a {\em resolved} photon. $x_\gamma$ is the fraction of the
photon's longitudinal momentum carried by the resolved gluon. Diagrams~(a) and (b) show the
single-Pomeron-exchange and the multi-Pomeron-exchange contributions, respectively. Similar diagrams apply to
diffractive dijet production in $pp$ collisions. \label{fig:2}}
\end{center}
\end{figure}
for dijet production) play an important role~\cite{A2,A3,A6}. It was demonstrated by the CDF
Collaboration~\cite{A4} that partonic distributions in the Pomeron (extracted from analyses of diffractive
production in DIS corresponding to the diagram of Fig.~1) lead, according to a na\"{\i}ve factorization
prescription (Fig.~2(a)), to a cross section which is about a factor of ten larger than the experimental one.
However, when multi-Pomeron $t$-channel exchange diagrams (Fig.~2(b)) are included using the framework of the
Reggeon diagram technique~\cite{A5}, which takes into account $s$-channel unitarity, the discrepancy disappears
and a good description of the CDF data is obtained~\cite{A6}. It is informative to extend the analysis to other
diffractive processes. We showed recently~\cite{A7} that the apparent breaking of QCD factorization in
double-Pomeron dijet production is also consistent with the same multi-Pomeron exchange model.

Here we address the question of whether such hard QCD factorization breaking takes place in photoproduction at
very small values of Bjorken $x$ at HERA. Experimental data for the diffractive photoproduction of dijets have
been obtained recently at HERA~\cite{ZEUS98,A8,A9}. We first comment on the interpretation of the recent H1 data
\cite{A8,A9}, and emphasize that the existing treatment does not lead to unique conclusions. We will consider a
simple analysis of experimental data, which also includes information on dijet inclusive photoproduction.
Theoretical predictions for cross section ratios will be given.

The second part of the paper is devoted to the important problem of unitarity effects (or multi-Pomeron exchanges)
in the lower part of the diagram of Fig.~1. It will be demonstrated that they are already crucial for the
distributions of gluons in the domain $x\lapproxeq10^{-4}$, almost independent of $Q^2$. The relation of these
effects to the `saturation' of partonic distributions is discussed.

\section{Diffractive DIS- and photo-production of dijets}

Here we consider the diffractive production of dijets by real and virtual photons in more detail. Besides the
diagram of Fig.~2(a), there is a large contribution to these processes from the diagram of Fig.~3,
\begin{figure}
\begin{center}
\includegraphics[height=7cm]{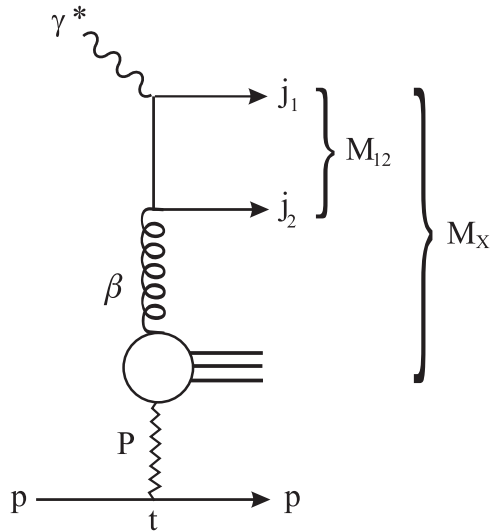}
\caption{Diffractive dijet production at HERA via a {\em direct} photon interaction. \label{fig:3}}
\end{center}
\end{figure}
which describes dijet production by photon--gluon fusion. This is usually called the `direct' contribution, while
Fig.~2(a) is known as the `resolved' contribution. For large $Q^2$ the effect of the rescattering diagrams of
Fig.~2(b) is expected to be small, and the cross section of diffractive dijet production can be written as a sum
of two terms
\be d\sigma^{jj}_D = d\sigma^{jj({\rm dir})}_D +  d\sigma^{jj({\rm res})}_D \label{eq:anne} \ee
with
\be  d\sigma^{jj({\rm dir})}_D = \int dt\int_{x_P^{\rm min}}^{x_P^{\rm max}} dx_P F_P(x_P,t)\beta  f_P^g(\beta,
\mu^2)\sigma_{\gamma g}^{j_1j_2}(E_{1T},E_{2T},M_{12}^2,\dots)\, d^n\tau \label{eq:betty} \ee
\be  d\sigma^{jj({\rm res})}_D = \int dt\int_{x_P^{\rm min}}^{x_P^{\rm max}} dx_P F_P(x_P,t)\beta  f_P^g(\beta,
\mu^2) x_\gamma f_{\gamma^*}^i(x_\gamma,\mu^2) \sigma_{i g}^{j_1j_2}(E_{1T},E_{2T},M_{12}^2,\dots)\, d^n\tau
\label{eq:cath} \ee
where $x_P = (Q^2 + M_X^2)/(Q^2 + W^2)$ is the fraction of the longitudinal momentum of the proton carried by the
Pomeron, and $\beta = (Q^2 + M_{12}^2)/(Q^2 + M_X^2)$ is the fraction of the Pomeron momentum carried by the
gluon. Here, $W$ is the total c.m. energy of the $\gamma p$ system, $M_X$ is the mass of the
diffractively-produced system and $M_{12}$ is the dijet mass. Also, $x_g = x_P\beta = (Q^2 + M_{12}^2)/(Q^2 +
W^2)$ is the fraction of the proton momentum carried by the gluon. Finally, $\tau$ denotes all the variables that
characterize the observed dijet system.

In \eq{cath}, $f_{\gamma^*}^i(x_\gamma,\mu^2)$ is the distribution of parton $i$ in the virtual photon carrying a
fraction $x_\gamma$ of its momentum. The Pomeron flux factor, $F_P(x_P,t)$, and distribution of gluons in the
Pomeron, $f_P^g(\beta, \mu^2)$, were determined from the analysis of data for the inclusive diffractive
dissociation of a virtual photon~\cite{A11,A12}. We take the scale $\mu^2 = Q^2 + \frac{1}{4}(E_{1T} + E_{2T})^2$.
Until recently there was a rather large uncertainty in the determination of the gluonic content of the Pomeron.
This uncertainty has been reduced in a recent LO and NLO analysis by the H1 Collaboration~\cite{A12}, which we
will use in the following.

The experimental situation concerning evidence for factorization breaking in the diffractive photoproduction of
dijets is far from clear. The current status can be summarized as follows. It was shown in Ref.~\cite{A8} that the
distribution of gluons from the new H1 analysis~\cite{A12} gives a prediction for the cross section of dijet
production in DIS, which is about 30\% below the experimental results. However, this prediction was obtained in a
LO QCD calculation, and NLO corrections can lead to an increase in the cross section, as was demonstrated for
inclusive dijet production~\cite{A13}. These effects are especially important in the small $x_\gamma$ region where
the resolved contribution mimics higher-order effects. On the other hand, the new analysis predicts the cross
section for the photoproduction of dijets to be about 30\% above experimental data. In this case, higher order
corrections can also be large and, moreover, the resolved contribution of Fig.~2(a), which is not well determined
at present, is rather substantial (especially at small $x_\gamma$). Thus, in our opinion, in this situation it is
difficult to draw a definite conclusion on the role of multi-Pomeron effects in $\gamma P$-dijet production. In
particular, the conclusion of Refs.~\cite{A9,A10} that there is an extra suppression of the $\gamma P$ interaction
by a factor $1.8\pm0.45$, which is independent of $x_\gamma$, should be treated with caution.

In order to clarify the situation we use a method, similar to that proposed by the CDF group for $\pp$
collisions~\cite{A4}, which includes extra information on totally inclusive dijet production. The last process is
described by the diagrams shown in Fig.~4.
\begin{figure}
\begin{center}
\includegraphics[height=5cm]{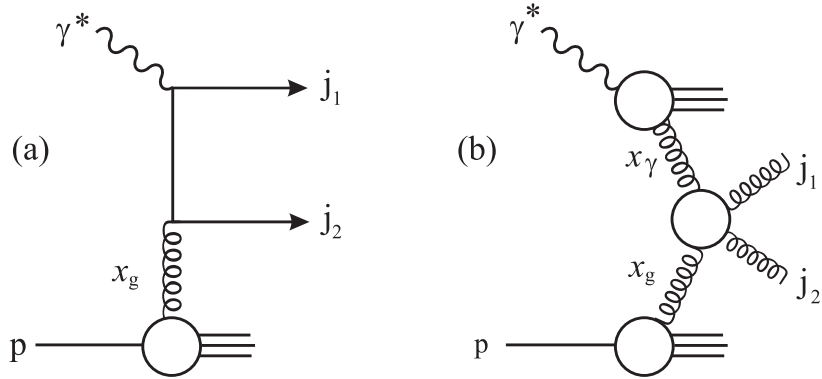}
\caption{Schematic diagrams for inclusive dijet production at HERA. This process is compared in the text to
diffractive dijet production. Diagrams (a) and (b) show the direct and resolved photon contributions respectively.
\label{fig:4}}
\end{center}
\end{figure}
The upper parts of these diagrams have the same structure as the diagrams of Fig.~2(a) and Fig.~3 and differ only
in the lower parts, which now depend on the distribution of gluons\footnote{For very small $x_g$, which is
relevant to the present analysis, gluons give the dominant contribution.} in the proton, rather than those in the
Pomeron. As a consequence, most quantities cancel in the ratio of diffractive and inclusive dijet production.
Indeed, we obtain
\be R \equiv \frac{d\sigma_D^{jj}(\tau)}{d\sigma^{jj}(\tau)} = \frac{\tilde
F_P^g(x_g,\mu^2)}{x_gf_p^g(x_g,\mu^2)}\,, \label{eq:ethel} \ee
where the distribution of gluons in the diffractive process (containing a rapidity gap) is given by
\be \tilde F_P^g(x_g,\mu^2) = \int dt \int_{x_P^{\rm min}}^{x_P^{\rm max}} dx_P\,F_P(x_P,t)f_P^g(\beta,\mu^2),
\label{eq:frances} \ee
with $x_P^{\rm min} = x_g$ ($\beta = 1$) and where $x_P^{\rm max}$ is usually chosen to correspond to the
experimental requirement on the rapidity gap. Note that, although $R$ does not depend explicitly on the variable
$x_\gamma$, this dependence appears due to kinematic correlations between the different variables. For example,
since $M_{12}^2$ is strongly peaked near the value related to the experimental lower cutoff, an $x_\gamma$
dependence arises via $x_\gamma x_g W^2 = M_{12}^2$.

To predict the ratio $R$ of the experimental cross sections, \eq{ethel}, we need to choose the input for
$F_P(x_P,t)$ and the distribution of gluons in the Pomeron. Here we used a factorized expression in terms of the
variables $x_P$ and $\beta$, which corresponds to the exchange of a single Pomeron pole $P$, and which was used in
analysis of experimental data\footnote{For not too small values of $x_P\sim 0.1$, the contributions of secondary
Reggeons can be important, and are usually included in the data analysis~\cite{A11,A12}.}. In general, we can have
more complicated, unfactorizable forms (see Fig.~5)
\begin{figure}
\begin{center}
\includegraphics[height=7cm]{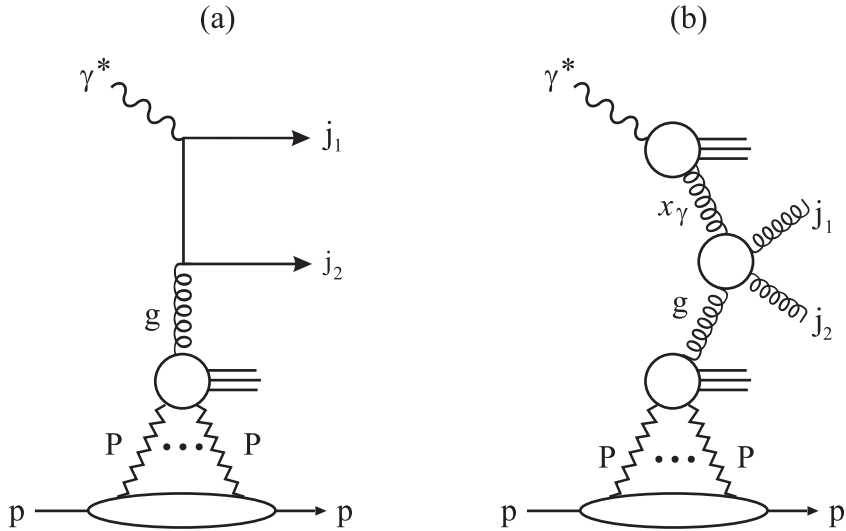}
\caption{Multi-Pomeron contributions to inclusive dijet production. \label{fig:5}}
\end{center}
\end{figure}
but the same analysis of the ratio $R$ still holds.

The theoretical prediction for the ratio $R$ is shown by the continuous curve in Fig.~6.
\begin{figure}
\begin{center}
\includegraphics[height=10cm]{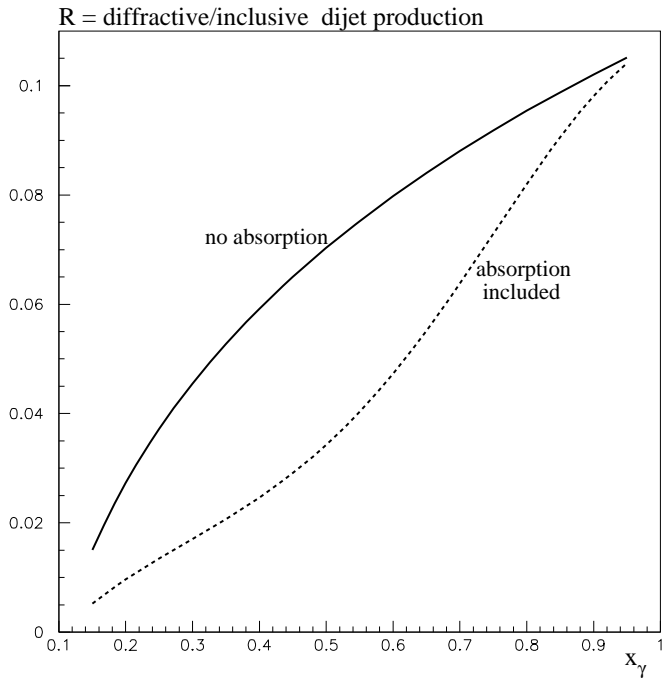}
\caption{The predictions for the ratio, $R$, of diffractive and inclusive dijet photoproduction at HERA, of
\eq{ethel}, as a function of $x_\gamma$. The curves have been calculated using the Pomeron flux and gluon
distribution in the Pomeron of Refs.~\cite{A12}, and correspond to a $\gamma$-proton c.m. energy $W=205$~GeV,
dijet mass $M_{12} = 12$~GeV, $x_P^{\rm max} = 0.03$ and scale $\mu^2=15$~GeV$^2$. For the gluon distribution in
the proton we conservatively use that of CTEQ6M partons~\cite{CTEQ6}. The use of the MRST2001 or
MRST2002~\cite{A14} gluons gives a value of $R$ which is a bit larger. The predictions based on single-Pomeron and
multi-Pomeron exchange are shown as continuous and dashed curves respectively. The ratio $R$ of the high $Q^2$
processes is not expected to have absorptive corrections, and hence should follow the continuous curve.
\label{fig:6}}
\end{center}
\end{figure}
It is calculated using $F_P$ and $f_P^g$ given in Ref.~\cite{A12} and $f_p^g$ from Ref.~\cite{CTEQ6}, together
with the experimental cuts of the H1 dijet experiment~\cite{A9}. As the result is sensitive to experimental cuts
it would be desirable to repeat the calculation of $R$ using a Monte Carlo with the exact data cuts.

So far we have considered the large $Q^2$ region, where the contributions of the diagrams of Fig.~2(b), with
unitarity corrections, are expected to be small. For small values of $Q^2$ (or real photons) these corrections
should be taken into account. However these corrections influence mainly the production of dijets by {\em
resolved} photons, which takes place mostly at small and moderate values of $x_\gamma$. For {\em direct}
production, via Fig.~3, the rescattering corrections are small. It is worthwhile to recall the reason. First, the
cross section of dijet rescattering is proportional to $1/E_T^2$ and therefore the contribution of
Fig.~\ref{fig:2}(b) is negligible. Next, it is important to note that the parton distributions of the Pomeron were
measured in {\em direct} inclusive processes such as those shown in Figs.~3 and 5(a). Thus they effectively
include the multi-Pomeron effects shown in Fig.~5, as well as the multi-Pomeron counterpart to Fig.~2(b) for the
`direct' process. The direct process is thus not suppressed by any additional rescattering corrections, as all
such effects are already embodied in the normalisation of the effective Pomeron structure functions.

These corrections, which are sometimes called `the survival probability of rapidity gaps' or `the screening
corrections from the underlying events', suppress the cross section for hard diffractive processes. A two-channel
eikonal model~\cite{A15}, with parameters tuned to fit all soft $pp$ (and $p\bar p$) data, has been used to
predict the suppression factors for various hard diffractive processes. This model was used for the description of
the diffractive production of jets in $p\bar p$ interactions~\cite{A6}. Here we apply the model to the
photoproduction of dijets\footnote{This model was applied to the photoproduction of jets separated by a large
rapidity gap in Ref.~\cite{MMR}.}. To describe the photon--proton interaction we use the generalized vector
dominance model. We tune the $\rho$-meson Pomeron vertex so that, at $W=200$~GeV, $\sigma^{\rm tot}(\rho p) =
34$~mb and the slope of the cross section for diffractive $\rho$ photoproduction is $B=11.3$~GeV$^{-2}$, to be
consistent with the HERA data~\cite{elastic}. The parameter $\gamma$ in the two-channel eikonal (eq.~(33) of
Ref.~\cite{A15}) was taken to be $\gamma=0.6$ to account for the large probability of $\rho$ meson excitation in
comparison with that for the proton.

In the {\em ideal} theoretical limit, using the above model we find that the difference between the predictions
for $R$ for DIS and for photoproduction is a common suppression factor of 0.34 for photoproduction for all
$x_\gamma$, except for the `direct' photon contribution at $x_\gamma=1$ which has no suppression. However, in {\em
reality}, the `direct' contribution is smeared by the experimental resolution and uncertainties connected with the
jet finding algorithms. In an attempt to account for the smearing, we assume that the direct contribution is of
the Gaussian form $\exp(-6(1-x_\gamma)^2)$, chosen to agree with the observations of Fig.~4 of Ref.~\cite{A9}.

With this smearing factor, and using the same experimental cuts for photoproduction as for the large $Q^2$ DIS
data, we predict the suppression corresponding to the dashed curve in Fig.~\ref{fig:6}. For small
$x_\gamma\lapproxeq0.3$, away from the smearing effects, we see the 0.34 suppression factor. That is the dashed
``photoproduction'' curve lies a factor 3 below the ``high $Q^2$'' continuous curve, which does not suffer
unitarity corrections.

We believe that this method of analysis is simple, informative and convenient from both the theoretical and
experimental points of view, as most of the theoretical uncertainties (as well as some experimental systematics)
cancel in the ratio $R$. Moreover, it can be used in other situations, such as, for example, the diffractive
production of charm~\cite{A16,A17}.

\section{Unitarization of the gluon distribution at small $x$}

Now we consider some problems with existing parameterizations of the quantity $\tilde F_P^g(x,\mu^2)$ of
\eq{frances}. By definition the quantity $R$ is less than unity. However, for $\alpha_P(0) = 1.173$ in the Pomeron
flux factor and with the existing expressions for $f_P^g(\beta,\mu^2)$~\cite{A12}, the function $R$ quickly
increases as $x\to 0$. This can be seen from Fig.~7,
\begin{figure}[!ht]
\begin{center}
\includegraphics[height=10cm]{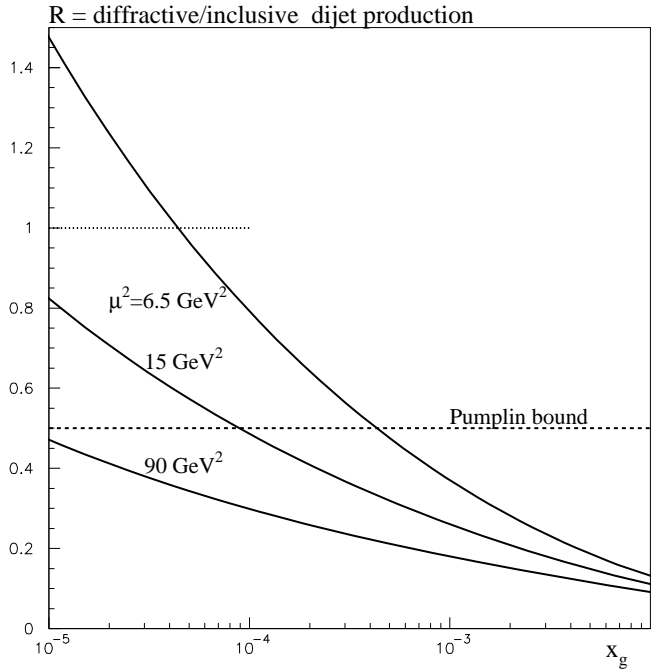}
\caption{Predictions for the ratio $R$ of diffractive and inclusive dijet production, of \eq{ethel}, shown as a
function of $x_g$ for different scales $\mu^2$, with $x_P^{\rm max}=0.1$. Absorptive corrections are neglected.
\label{fig:7}}
\end{center}
\end{figure}
which shows the ratio $R$ for different values of $\mu^2$. Indeed we see that $R$ approaches, or even exceeds,
unity in the region\footnote{Of course, in this region the curves are particularly sensitive to the choice of
input for the Pomeron flux, the Pomeron and proton gluon densities, and the value of $x_P^{\rm max}$, but all
reasonable choices indicate a violation of unitarity similar to that shown in Fig.~\ref{fig:7}.} of
$10^{-4}\lapproxeq x_g \lapproxeq 10^{-5}$ for quite a range of virtualities $\mu^2$. This means that the
application of the single Regge pole approximation in the small $x$ region is not valid and leads to a violation
of unitarity. The diagrams of Fig.~5, which take into account multi-Pomeron exchanges, must be included (as, for
example, has been done in Ref.~\cite{A18}) in order to ensure that $R<1$, and to restore unitarity\footnote{See
also the discussion of this problem in Ref.~\cite{A19}.}.

Comparing the diagrams of Figs.~\ref{fig:5} and \ref{fig:4}, we may say that the value of $R$ of (\ref{eq:ethel})
represents the ratio of the cross section for gluon diffractive dissociation (the lower part of Fig.~\ref{fig:5})
to the total gluon--proton cross section. In analogy with hadronic interactions, it is reasonable to believe that
diffractive processes are less than one half of the total gluon--proton cross section~\cite{PUMP}, and thus that
$R\leq\frac{1}{2}$. From Fig.~7 we see that, for gluons, this bound is already exceeded for $x_g\sim10^{-4}$.
Moreover, the violation of unitarity appears in this region of $x_g$ over a large interval of $\mu^2$. This is
related to the fact that the diffractive production of states with small $\beta$ is not a high-twist effect. Note
that here we discuss the absorptive effect caused by the rescattering of intermediate partons (described by the
multi-Pomeron exchange contributions shown in Fig.~\ref{fig:5}), and not the rescattering of the fast constituents
of the photon. The rescattering of intermediate partons takes place at scales much smaller than the hard scale
$\mu^2$. The main origin of the large values of $R$ (which indicate the violation of unitarity) is the power-like
growth ($x^{1-\alpha_P(0)}$) of the single-Pomeron-exchange amplitude. This growth must be tamed. As can be seen
from Fig.~\ref{fig:7}, this effect is much more important in inelastic diffraction than in the total cross section
or pure inclusive processes. In inclusive processes the absorptive corrections are small due to AGK
cancellations~\cite{AGK}\footnote{Indeed soft rescattering does not alter the distribution of high $E_T$ jets in
inclusive processes, but fills in and destroys the rapidity gap in diffractive reactions.}. In the total cross
section, or in inclusive DIS discussed here, part of the absorptive effect is mimicked by the input gluon
distribution.

These observations may be important for understanding the problem of `saturation' in heavy-ion collisions, and
also in diffractive charm production at very small $x$.

\section{Conclusions}

The recent observation of diffractive dijet photoproduction, combined with the measurements of diffractive dijet
DIS production, offers a unique opportunity to probe the special features of diffractive dynamics. We have
emphasized that a good way to study the effects of factorization breaking due to rescattering (or, in other words,
the suppression caused by absorptive corrections) is to take the ratio of the diffractive process to the
corresponding inclusive production process, that is
\be R = \frac{\sigma({\rm diffractive})}{\sigma({\rm inclusive})} \label{eq:gertrude} \ee
of (\ref{eq:ethel}). A comparison of the predictions for $R$ with the data, for both photoproduction and DIS,
should be very informative. Moreover many theoretical uncertainties, and some experimental uncertainties, cancel
in the ratio.

The high $Q^2$ diffractive process, unlike diffractive photoproduction, is not expected to suffer rescattering
corrections. Indeed in the ideal theoretical limit we predict
\be R({\rm photoprod.})\ \simeq\ 0.34\, R({\rm high}\ Q^2), \label{eq:holly} \ee
except at $x_\gamma=1$, where the ratios are expected to be equal since there should be very small rescattering
corrections to the `direct' photon contribution. In practice the experimental resolution and jet finding
algorithms have the effect of smearing the `direct' contribution. We estimated this effect and showed the
resulting predictions for $R$ in Fig.~\ref{fig:6}. The 0.34 suppression of diffractive photoproduction due to
absorptive corrections is clearly evident for small $x_\gamma\lapproxeq0.3$, where smearing effects are
negligible. Since the predictions for $R$ depend on the experimental cuts (in particular on the value of $x_P^{\rm
max}$) the curves should be recalculated to match the conditions of the experiment, including the smearing of the
$x_\gamma$ distribution. However the ratio of ratios, $R({\rm photoprod.})/R({\rm high}\ Q^2)$, for
$x_\gamma\lapproxeq 0.3$ should be a reasonably stable prediction.

Clearly it will be interesting to measure $R$ as a function of $Q^2$. At large $Q^2$ the absorptive corrections
are expected to be negligible and the prediction is given by the factorization theorem, that is the continuous
curve in Fig.~\ref{fig:6}. However as $Q^2$ decreases we would expect the prediction for $R$ to tend gradually
towards the dashed photoproduction curve. Recall that for $x_\gamma\gapproxeq0.5$ the values of $R$ have an
additional uncertainty due to the contamination of the `direct' photon contribution.

The above behaviour of $R$ is also expected for diffractive charm production. However, in this case, the relative
contribution of the `direct' photon process (where the absorptive effect is expected to be small) is enhanced by a
colour factor.

The value of $R$ is proportional to the effective number of gluons coming from the Pomeron, and may be considered
as the ratio of gluon diffractive dissociation to the total gluon--proton cross section. At relatively low scales,
we see from Fig.~\ref{fig:7}, the ratio in the absence of absorptive corrections exceeds the Pumplin bound,
$R<\frac{1}{2}$, already at $x_g\sim10^{-4}$. Thus it reveals the need for absorptive corrections\footnote{The
multi-Pomeron contributions in Fig.~\ref{fig:5}.} in diffractive processes already at HERA energies.



\section*{Acknowledgements}

We thank Sebastian Sch\"atzel for information. ABK and MGR would like to thank the IPPP at the University of
Durham for hospitality. This work was supported by the UK Particle Physics and Astronomy Research Council, by
grants INTAS 00-00366, RFBR 00-15-96786, 01-02-17383 and 01-02-17095, and by the Federal Program of the Russian
Ministry of Industry, Science and Technology 40.052.1.1.1112 and SS-1124.2003.2.





\end{document}